\documentclass[showpacs,reprint,twocolumn,preprintnumbers,amsmath,amssymb,aps,superscriptaddress]{revtex4-1}

\usepackage{graphicx}
\usepackage{dcolumn}
\usepackage{bm}
\usepackage{subfigure}
\usepackage{color}
\usepackage{times}

\renewcommand{\eqref}[1]{(\ref{#1})}


\begin{document}

\title{Phase separation and pattern formation in a binary Bose-Einstein condensate}
\author{Jacopo Sabbatini}
\email{sabbatini@physics.uq.edu.au}
\affiliation{The University of Queensland, School of Mathematics and Physics, Qld 4072, Australia}
\author{Wojciech H. Zurek}
\affiliation{Theory Division, Los Alamos National Laboratory, Los Alamos, New Mexico 87545, USA}
\author{Matthew J. Davis}
\affiliation{The University of Queensland, School of Mathematics and Physics, Qld 4072, Australia}

\date{\today}

\begin{abstract}
The miscibility-immiscibility phase transition in binary Bose-Einstein condensates (BECs) can be controlled by a coupling between the two components. Here we propose a new scheme that uses coupling-induced pattern formation to test the Kibble-Zurek mechanism (KZM) of topological-defect formation in a quantum phase transition.  For a binary BEC in a ring trap we find that the number of domains forming the pattern scales as a function of the coupling quench rate with an exponent as predicted by the KZM.  For a binary BEC in an elongated harmonic trap we find a different scaling law due to the transition being spatially inhomogeneous.   We perform a ``quantum simulation'' of the harmonically trapped system in a ring trap to verify the scaling exponent.
\end{abstract}

\maketitle

The formation of topological defects in symmetry-breaking phase transitions is a universal phenomenon relevant to many fields of physics, from cosmology \cite{texturecmb} to condensed matter \cite{chuang,mcclintock94,ruutu,bauerle,dodd98,monacorivers2006}. 
The Kibble-Zurek mechanism (KZM) \cite{Kibble:1976vm,kibble80,zurek85,Zurek:APP1993,Zurek:1996uj} is a theory relating the density of defects in the broken symmetry phase to the timescale of the transition.  The scaling law predicted by the KZM has been demonstrated in simulations of phase transition dynamics \cite{zurek1997,zurek2002,retzkerPRL2010,Damski:2007ek,ueda2007,uedaKZspinor},
and has motivated many experiments \cite{mcclintock94,dodd98,ruutu,bauerle,monacorivers2006,carmi,maniv,Kirtley:2003ke,Monaco:2009hg,Golubchik:2010ba}.  However, the KZM scaling of defect density with the  quench rate has not been verified in the laboratory.

An experiment aiming to observe the Kibble-Zurek (KZ) scaling of defect production must have  good control of the progress of the system through the phase transition.
The extraordinary degree of flexibility and control available in ultra-cold gas experiments
makes them leading candidates for the accurate testing of the KZM predictions.
For example, the spontaneous formation of vortices in a Bose-Einstein condensate (BEC) following the evaporative cooling of a single-component thermal Bose gas was suggested as a candidate system for studying the KZM \cite{Anglin1999}.  Recently two experiments have observed this phenomenon, although they were not suited for testing the KZM \cite{matt,hall2010}.  
As Bose-Einstein condensation is a classical (thermal) phase transition, the formulation of a KZ scenario for this system requires the control of thermodynamic parameters such as the temperature $T$ and/or the chemical potential $\mu$.  This is difficult for an isolated quantum system  such as an ultra-cold gas, as the system thermalises under its own dynamics rather than through a coupling to a thermal reservoir.

The KZM, however, has been adapted  to describe defect formation in quantum phase transitions \cite{Zurek:2005cu,Dziarmaga:2005jq,Polkovnikov:2005gr}, in which a  Hamiltonian parameter is ramped through a quantum critical point.  
Hamiltonian quenches  are relatively straightforward to control in an isolated quantum system, and thus suggests that quantum  phase transitions are strong candidates for quantitative KZM studies with ultra-cold gases. An important experiment on this topic was  a study of the dynamics of  a spin-1 BEC following a magnetic field quench~\cite{sadler06}, resulting in the formation of topological defects such as spin-vortices and polar-core vortices~\cite{sadler06}.
This experiment was analysed in the context of the KZM in \cite{ueda2007,uedaKZspinor,Damski:2007ek}, but the KZ scenario in this system is complicated by the unclear role of dipolar interactions \cite{Vengalattore:2010hz} and the difficulties involved in imaging and counting the topological defects~\cite{sadler06}.  A quantum quench from the Mott insulator to superfluid state in a optical lattice was recently reported in \cite{deMarco}, but did not demonstrate KZ scaling.

In this Letter we formulate a straightforward, experimentally realistic Kibble-Zurek scenario for a coupling-induced miscibility-immiscibility quantum phase transition in a binary Bose-Einstein condensate \cite{miscibility}. The observable is the number of domains formed in the immiscible phase. The  domain-walls are stable and long-lived in a ring trap, and may be easily detected using absorption imaging \cite{counting_note,Hamner:2011bq}.  
The transition is achieved by reducing the coupling between the two components provided by a microwave or laser field, ensuring precise control of the quench.

We find that a coupling quench of a binary BEC in a ring trap confirms the KZ theory predictions for the resulting number of defects.  However, we find that the same transition in  a quasi-1D harmonically trapped BEC yields a different scaling exponent  for the defect density as compared to the ring BEC and the KZ prediction. We attribute the difference to the motion and decay of the defects. To confirm this we design a spatially dependent quench of the Hamiltonian parameters in the ring trap such that the experiment has the same characteristics as the spatially inhomogeneous phase transition occurring in a harmonically trapped binary BEC.

The physics of binary BECs has been extensively studied in previous work \cite{binarybec_ho,binarybec_bigelow}.
Here we consider a binary BEC formed by a single atomic species with two hyperfine states ($i = 1,2$) that are linearly coupled with time-dependent amplitude $\Omega(t)$, such as may be achieved with resonant two-photon microwave coupling \cite{Hall:1998fa}.  We restrict ourselves to an elongated system where the spin dynamics are confined to the $x$ dimension.  We integrate out the transverse degrees of freedom (assumed to be harmonically trapped with  frequency $\omega_\perp$), and utilise the rotating-wave approximation (RWA)  to give the one-dimensional Hamiltonian
\begin{eqnarray}\label{eq:hamiltonian}
 \hat{H}& =& \int dx\left[\sum_{i=1,2}\left(\hat{\psi}_{i}^{\dagger}H_0 \hat{\psi}_{i} \nonumber
  + {\frac{g_{ii}}{2}}  \hat{\psi}_{i}^{\dagger}\hat{\psi}_{i}^{\dagger}\hat{\psi}_{i}\hat{\psi}_{i}\right)
+\hbar\delta\hat{\psi}_{2}^{\dagger}\hat{\psi}_{2} \right.\\
 &&+ \left. g_{12}\hat{\psi}_{1}^{\dagger}\hat{\psi}_{2}^{\dagger}\hat{\psi}_{2}\hat{\psi}_{1}-\hbar\Omega(t)(\hat{\psi}_{1}^{\dagger}\hat{\psi}_{2}+{\rm h.c.})\right] ,
\end{eqnarray}
where $H_0 = -\frac{\hbar^2}{2m} \frac{d^2}{dx^2} + V(x)$ is the single-particle Hamiltonian, and $\hat{\psi}_{i} \equiv \hat{\psi}_{i}(x)$ is the Bose field operator for component $i$. The detuning of the transition is $\hbar \delta$, which we subsequently  set to zero. The nature of the ground state of the uncoupled system with $\Omega(t) = 0$ is determined by  the parameter \cite{binarybec_ho} $\Delta = g_{11}g_{22}/g_{12}^2$, where $g_{12}$ ($g_{ii}$) is the 1D inter-species  (intra-species) interaction constant, with $g_{ij} = 2 \hbar^2 a_{ij} / (m a_\perp)$, $a_\perp=(\hbar/m\omega_\perp)^{1/2}$ and $a_{ij}$ are the scattering lengths.  The Gross-Pitaevskii  ground state \cite{pethicksmith} of Eq.~\eqref{eq:hamiltonian} with $V(x) = \Omega(t) = 0$ is found to be miscible for  $\Delta>1$ \cite{binarybec_ho}.

Here we consider a system with $\Delta<1$ that is phase-separated  for $\Omega(t)=0$, but  becomes miscible above a density dependent critical coupling strength $\Omega_{\rm cr}$ \cite{miscibility}.  If we begin an experiment in the ground state with $\Omega(0) > \Omega_{\rm cr}$, the system will phase separate as  $\Omega(t)$ is ramped to zero. If this occurs sufficiently quickly, the system cannot adiabatically follow its ground state, and a spatially random pattern of domains consisting entirely of atoms in either state 1 or 2 will form.
Our proposal will allow for a test of the KZM by controlling the rate at which $\Omega(t)$ is reduced to zero, and counting the final number of domains.  

\begin{figure}
\includegraphics[width=0.46\textwidth]{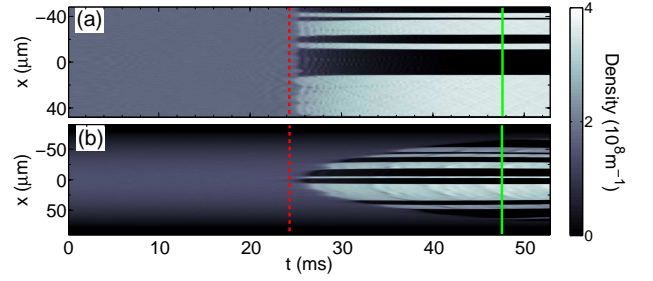}
\caption{(Color online) Examples of domain formation in the density of one hyperfine  component for (a) a uniform system with $\tau_{\rm Q}=63$ ms and $N = 10^5$ atoms and (b) a trapped system with similar parameters. The red dashed vertical lines indicate the time where $\Omega(t)=\Omega_{\rm cr}$, while  the solid green vertical lines indicate the end of the quench $\Omega(t) = 0$.  The density of the second component (not shown) is complementary to the density of the first component.}
\label{fig:uniex}
\end{figure}

We first consider a periodic uniform system  [$V(x)=0$] of length $L$ as realised by a binary BEC in a ring trap \cite{Henderson:2009eo,Gupta:2005ed,Ramanathan:2011bi}.  Within the mean-field approximation, the energy spectrum of the Hamiltonian \eqref{eq:hamiltonian}  has a gap $E_{\rm gap}\propto\hbar\sqrt{\Omega(t)[\Omega(t)-\Omega_{\rm cr}]}$  in the miscible regime \cite{tommasini,lee09}, where $\hbar\Omega_{\rm cr} = 2(g_{12}^2 - g_{11}g_{22})\rho/(g_{11}+g_{22}+2g_{12})$ is the  coupling strength that defines the quantum critical point  and $\rho$ is the linear atom density  \cite{note_omegacr}.
To formulate the  Kibble-Zurek scenario for this system, we define a control parameter $\epsilon(t)$ measuring the distance of the system from the critical point
\begin{equation}\label{eq:controlparameter}
\epsilon(t) = 1 - \Omega(t)/\Omega_{\rm cr}.
\end{equation}
In the thermodynamic limit, when $\epsilon(t)\rightarrow0$ the correlation length $\xi$ and the relaxation time $\tau$ diverge as $\xi = \xi_{0}/|\epsilon|^{\nu}$, $\tau = \tau_{0}/|\epsilon|^{\nu z}$, where the spatial and dynamical critical exponents ($\nu$ and $z$ respectively) depend only on the universality class of the transition.  Due to the divergence of the relaxation time $\tau$ as $\epsilon \rightarrow 0$, we expect the system evolution to become non-adiabatic when the time needed to adjust to an external change is equal to the timescale at which the coupling changes \cite{zurek85}
\begin{equation}\label{eq:freezingtime}
\tau(\hat{t}) = \epsilon(\hat{t})/\dot{\epsilon}(\hat{t}),
\end{equation}
which defines the freezing time, $\hat{t}$.
Choosing a linear ramp $\Omega(t) = \mathrm{max}\left[0,2\Omega_{\rm cr}\left(1-{t}/{\tau_{\rm Q}}\right)\right]$ for the coupling, where $\tau_{\rm Q}$ is the quench time, we find that the correlation length at the freezing time is $\xi(\hat{t}) = \xi_{0} (\tau_{\rm Q}/\tau_{0})^{{\nu}/({1+\nu z})}$.   For our system the mean-field critical exponents are  $\nu=1/2$ and $z=1$ \cite{lee09}, with $\xi_{0}=\xi_{s}/\sqrt{2}$ where  the spin healing length $\xi_{\rm s}=\hbar/\sqrt{2m \rho g_{\rm s}}$, and $\tau_{0}=\hbar/2g_{\rm s}\rho$ with $g_{\rm s}=(g_{11}+g_{22}-2g_{12})/2$.  The  prediction for the mean number of domains $N_{\rm d}$ at the end of the quench is
\begin{equation}\label{eq:KZscaling}
N_{\rm d} = L /  \xi(\hat{t})  = (L/\xi_{0}) (\tau_{0}/\tau_{\rm Q})^{1/3}.
\end{equation}
\indent To test this prediction, we perform quantum dynamical simulations of the proposed experiment for a range of quench times $\tau_{\rm Q}$ using the truncated Wigner method \cite{mattreview}.  This is equivalent to adding a half quantum of noise per mode to the initial mean-field  wave function, and then 
numerically solving the time-dependent Gross-Pitaevskii equation for the binary system  \cite{ICnote}.  
Quantum expectation values for the dynamics then can be calculated by averaging over an ensemble of trajectories. We interpret each trajectory as an individual experimental realisation \cite{mattreview}, and count the number of domains at the end of the evolution. A typical trajectory is shown in Fig.~\ref{fig:uniex}(a), where domain formation is observed  once $\Omega(t) < \Omega_{\rm cr}$. At times when $ 0 < \Omega(t) < \Omega_{\rm cr}$ the domains can drift and merge \cite{FollowUpPRA}, however once $\Omega(t)=0$   the pattern of defects is stable, allowing for their unambiguous counting \cite{counting_note}. 

\begin{figure}
\includegraphics[width=0.42\textwidth]{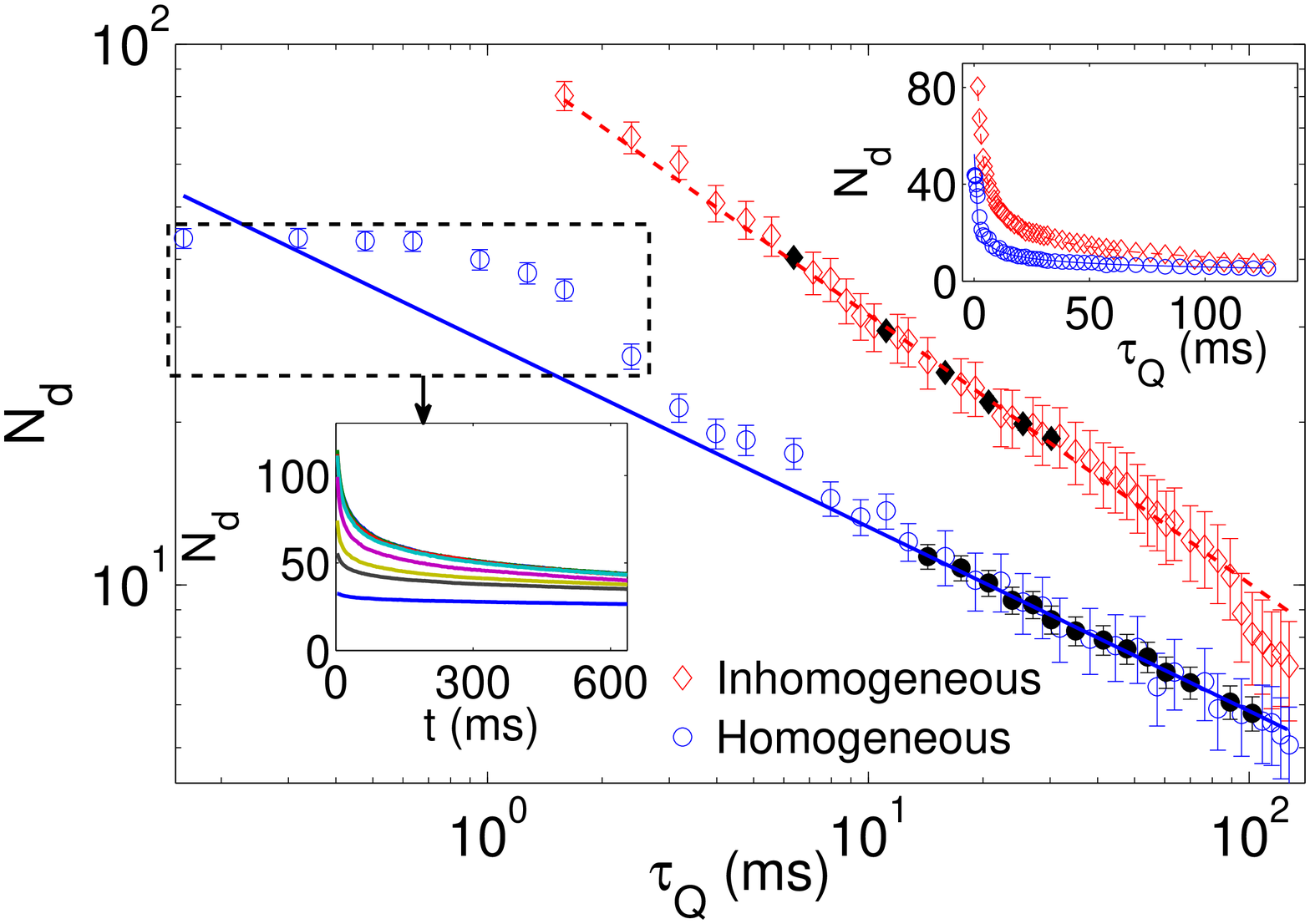}
\caption{(Color online) Mean number of domains formed $N_{\rm d}$ as a function of the quench time $\tau_{\rm Q}$ for a BEC in a ring trap.  Blue circles are for a spatially homogeneous quench, and the red diamonds are for a spatially inhomogeneous quench simulating a harmonically trapped system. The open symbols average over 100 trajectories, while the closed symbols average over 1000 trajectories and are used for fitting the scaling exponent $n$.   The error bars indicate the standard error.  The linear fits to the data yield a scaling exponent $n=0.341\pm 0.006$ for the homogeneous quench, and $n=0.497\pm0.015$ for the inhomogeneous quench.  Data for the inhomogeneous quench are shifted upwards by a multiplicative factor for clarity. The top right inset shows the same data on a linear scale. The bottom left inset shows the time evolution of the average number of domains for quenches with $\tau_Q \le 2$ ms, for which  the mean number of domains is still decreasing at $t= 600$ ms. 
}
\label{fig:powerlaw}
\end{figure}

For the ring system we choose parameters $N = 5 \times 10^4$ atoms, $L= 96$~$\mu$m, $a_{11} = a_{22} = a_{12}/2 = 1.325 $ nm, and $\omega_\perp/ 2\pi = 2$ kHz. With such parameters we have $\xi_{\rm s}=1.16$ $\mu$m, $\xi_{0}=0.82$ $\mu$m and $\tau_{0}=1.8$ ms. The spin healing length $\xi_{\rm s}$ is about two-thirds of the transverse system size, ensuring that the quasi-1D approximation is valid.  We simulate quench times  $\tau_{\rm Q}$ over three decades in the range  $[0.1,125]$~ms, and plot the mean number of domains $N_{\rm d}$ versus the quench time $\tau_{\rm Q}$ in Fig. \ref{fig:powerlaw}.  We fit a power law  $N_{\rm d} \propto \tau_{\rm Q}^{-n}$ to the data for $\tau_{\rm Q} \ge 2$ ms, and find $n = 0.341\pm 0.006$ in good agreement with the KZM prediction of $n = 1/3$.    Thus  a binary BEC in a ring trap is a strong candidate system for an experimental test of the KZM.

We note that our numerical results  deviate from the KZM prediction for rapid quenches with $\tau_{\rm Q} < 2$ ms.  For these quench times we find the number of domains is still decreasing at the end of our integration time  (see inset of Fig.~\ref{fig:powerlaw}) and thus the mean number of domains plotted in Fig.~\ref{fig:powerlaw} \emph{overestimates} the true final number.  Numerical instabilities, combined with the known limitations of the truncated Wigner method,  prevent us from further extending the integration time \cite{mattreview}.  The upper limit to the number of domains is apparent, being $N_{\rm d}^{\rm max}\approx L/\xi_{s} \approx 80$.

We now perform simulations of a quasi-1D harmonically trapped system [$V(x) = m \omega^2 x^2/2$] with $\omega/2\pi = 5$ Hz and other parameters as for the uniform ring trap.  For a system with nonuniform density $\rho(x)$  the critical coupling strength $ \Omega_{\rm cr}(x)$, and hence the control parameter $\epsilon_{\rm tr}(x,t),$ are spatially dependent and therefore the quantum phase transition is spatially inhomogeneous. The denser central region of the condensate enters the immiscible phase earlier, and a moving front separates it from the parts still in the miscible phase \cite{SeeSuppMat} as seen in Fig.~\ref{fig:uniex}(b).

\begin{figure}
\includegraphics[width=0.45\textwidth]{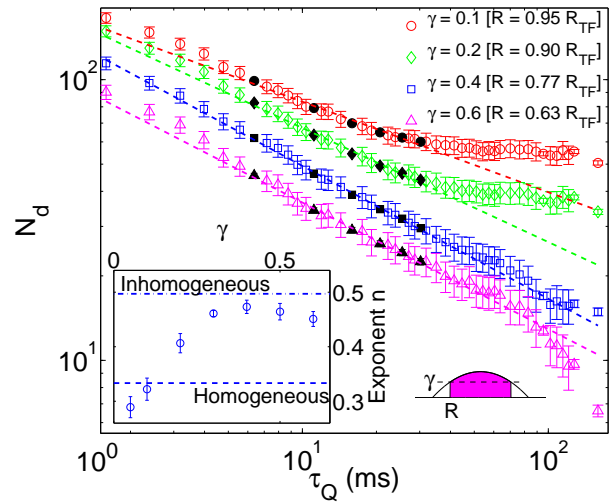}
\caption{(Color online) Mean number of domains formed $N_{\rm d}$ versus quench time $\tau_{\rm Q}$ for a harmonically trapped BEC.  The different colors (symbols) correspond to different domain-counting threshold densities $\gamma=\rho_{\rm cut}/\rho(0)$ (see text). Filled symbols represent ensembles of 1000 trajectories used for estimating the scaling exponent $n$. The left inset shows the value of  the exponent $n$ as a function of the density threshold $\gamma$. The dashed (dot-dashed) line indicates the value of $n$ obtained from a simulation of a homogeneous (inhomogeneous) coupling quench in a ring trap as in Fig.~\protect\ref{fig:powerlaw} (see  text). The shaded region in the right inset illustrates the counting region in the BEC density. The flattening of the domain number at long quench times for $\gamma=0.1$ and $\gamma=0.2$ is due to the unavoidable miscounting of density fluctuations as domains.}
\label{fig:thresholdscounts}
\end{figure}

The stochastic nature of the simulations means that it is difficult to distinguish domains from density variations near the edge of the condensate, and hence the counting of domains after the quench in the trapped system is problematic.
 To overcome this issue, we restrict our counting to the central region, where the initial density is greater than a threshold value $\rho_{\rm cut} = \gamma \rho(0) $.   Figure~\ref{fig:thresholdscounts} shows results  for the final number of domains as a function of quench time for a range of $\gamma$.  The inset of Fig.~\ref{fig:thresholdscounts} shows the scaling exponent resulting from the fitting of these curves to  $N_{\rm d} \propto \tau_{\rm Q}^{-n}$ as a function of $\gamma$.  The approximately constant value of $n\approx 0.47 $ for $\gamma > 0.3$ suggests that  variation of the scaling exponent with $\gamma$ is mostly due to the miscounting of domains.

To help understand the inhomogeneous phase transition, we  model a ``quantum simulation'' of the   spatially uniform quench in a harmonically trapped BEC by implementing a spatially dependent ramp of the coupling strength  in a ring BEC.
This is achieved by choosing $\Omega(x,t) = \mathrm{max}[0,2 \Omega_{\rm cr} (1-{t}/{\tau_{\rm Q}}) \rho(0)/\rho(x)]$ with $g=g_{11}=g_{22}=g_{12}/2$, where $\rho(x)$ is the ground state density of the harmonically trapped system we are modelling.  For such a quench the control parameter in the ring is spatially dependent and is identical to  $\epsilon_{\rm tr}(x,t)$ for the harmonically trapped BEC. To avoid the divergence of $\epsilon_{\rm tr}(x,t)$ where $\rho(x)\to 0$, we choose the circumference of the ring such that $L<2 R_{\rm TF}$, where $R_{\rm TF}$ is the Thomas-Fermi radius of the trapped system being simulated \cite{FollowUpPRA}.  We plot the number of domains versus quench time in  Fig.~\ref{fig:powerlaw}, and find clear evidence for  power-law scaling with an exponent $n = 0.497 \pm 0.015$.
This ``quantum simulation'' demonstrates that our result for the scaling exponent in the harmonic trap is robust, but we are unable to obtain this result analytically  \cite{FollowUpPRA}. 

We identify two effects potentially contributing to the increase in the scaling exponent in the harmonic trap compared to the ring trap.  First, during an inhomogeneous phase transition the  difference between the coupling and the critical coupling strength is spatially dependent, and this introduces a preferred direction for the movement of domains. The breaking of translation invariance leads to a larger annihilation rate of domains than in a homogeneous phase transition. During longer quenches domains have more  time to annihilate or escape the counting region, resulting in an increase of the observed scaling exponent \cite{FollowUpPRA}.

Second,  it has been proposed that when the front velocity $v_{\rm F}(x)$ is less than the local speed of sound $v_{\rm s}(x)$, the parts of the system in the symmetry-broken phase can influence the symmetry breaking in the region undergoing the transition, leading to a suppression of domain formation. This  was previously discussed  with regard to soliton formation in a thermally quenched Bose gas in Ref.~\cite{zureksoliton}.    
Below a certain $\tau_{Q}$ there is  a spatial region for which the velocity of the front is less than the speed of sound, and this increases in size with increasing  $\tau_{Q}$.  Thus slower quenches  result in a larger spatial region in which domain formation is suppressed, and this  contributes to a larger scaling exponent.

Finally, we consider the feasibility of the experiments we propose.   No pair of hyperfine states of $^{87}$Rb or $^{23}$Na naturally satisfies the immiscibility criterion with a spin healing length $\xi_{\rm s}$ sufficiently small to allow for the formation of multiple domains.  However, the two hyperfine states $|1\rangle=|F=1,m_{F}=+1\rangle$ and $|2\rangle=|F=2,m_{F}=-1\rangle$ of $^{87}$Rb have  $g_{11} \approx g_{22}$, and exhibit an interspecies Feshbach resonance that could be utilised to tune $g_{12}$  \cite{hamburg,Tojo:2010iz}.  We estimate that it is possible to attain $\Delta\approx 0.8$ while keeping inelastic losses sufficiently small to allow enough time to realise the proposed experiment. For a ring BEC of 5000 atoms with $\omega_\perp/2\pi = 2$ kHz and a circumference of 50 $\mu$m, $\Delta=0.8$ implies $\xi_{\rm s}= 0.4$ $\mu$m, $\tau_{0}= 351$ ms and $N_{\rm d}^{\rm max}\approx 50$. 
 
 The miscible-immiscible phase transition in binary condensates can also be controlled via the spin-orbit coupling of neutral atoms \cite{Lin:2011ba}. In this situation the phase transition to the immiscible state is achieved by ramping up the intensity of two slightly detuned lasers coupling two hyperfine levels of $|F=1\rangle$ of $^{87}$Rb. This scheme has the significant advantage of being able to reach the strongly immiscible phase without suffering from the inelastic atom losses  common near a Feshbach resonance \cite{Ketterle:1998if}. 
 
  For both schemes the stability of domains is ensured far from the transition, as each component acts as an effective potential for the other.   In the strongly immiscible regime $\Delta\lesssim 0.95$ the kinetic and thermal energy are not sufficient to overcome the barriers provided by the domain pattern.  

In conclusion, we have shown that the number of domains arising in  coupling-induced pattern formation in a  binary BEC in a ring trap scales as predicted by the Kibble-Zurek mechanism. Recent demonstrations of ring BECs \cite{Henderson:2009eo,Gupta:2005ed,Ramanathan:2011bi}, combined with the experimental feasibility of the scheme, make it  an excellent candidate for testing the Kibble-Zurek theory.  We have also verified that a scaling law exists for harmonically trapped BECs, allowing for a qualitative test of the KZM in this system.

The authors thank Bogdan Damski and Markus Oberthaler for useful discussions, and Adolfo del Campo, Tod Wright and Arnab Das for carefully reading the manuscript.  This research was supported by the Australian Research Council through the ARC Centre of Excellence for Quantum-Atom Optics, and Discovery Project DP1094025. We acknowledge the support of U.S. Department of Energy through the LANL/LDRD program.
\bibliographystyle{apsrev4-1}

\begin{thebibliography}{56}
\bibitem [{not()}]{note_omegacr}%
  \BibitemOpen
  \href@noop {} {}\bibinfo {note} {The value of $\Omega_{\rm cr}$ was obtained
  in the equal-particle-number approximation $N_1 = N_2 = N / 2$ with the
  linear density $\rho=N/2L$. We have checked that this expression also holds
  for the harmonically trapped case within the local density
  approximation.}\BibitemShut {Stop}%
\bibitem [{ICn()}]{ICnote}%
  \BibitemOpen
  \href@noop {} {}\bibinfo {note} {The truncated Wigner initial condition for
  our simulations is the mean-field ground state of Eq.~\eqref{eq:hamiltonian}
  for $\Omega(0)=2\Omega_{\rm cr}$,
  $(\psi_{1},\psi_{2})=(\sqrt{N/2L},\sqrt{N/2L})$, with the addition of complex
  Gaussian noise to the Bogoliubov modes of the system with a mean particle
  number of 1/2.}\BibitemShut {Stop}%
\end{thebibliography}

\end{document}


\title{Phase separation and pattern formation in a binary Bose-Einstein condensate --- \\ Supplementary material}
\author{Jacopo Sabbatini}
\affiliation{The University of Queensland, School of Mathematics and Physics, Qld 4072, Australia}
\author{Wojciech H. Zurek}
\affiliation{Theory Division, Los Alamos National Laboratory, Los Alamos, New Mexico 87545, USA}
\author{Matthew J. Davis}
\affiliation{The University of Queensland, School of Mathematics and Physics, Qld 4072, Australia}

\date{\today}

\maketitle

\section{Critical coupling}
The critical coupling for the homogeneous case $\Omega_{\mathrm{cr}}$ is determined for the case of $N_{1}=N_{2} = N/2$ with the linear density $\rho=N/2L$.  The result (reproduced from the main text) is 
\begin{equation}
\hbar\Omega_{\rm cr} = \frac{2(g_{12}^2 - g_{11}g_{22})\rho}{g_{11}+g_{22}+2g_{12}}.
\end{equation}
For the harmonically trapped system, we use the Thomas-Fermi approximation \cite{pethicksmith} for the ground state density  condensate in the miscible regime when  $\Omega(0) > \Omega_{\rm cr}$ 
\begin{equation}\label{eq:TFC}
\rho(x)=\frac{\mu-V(x)+\hbar\Omega(0)}{g_{12}+g},
\end{equation}
where we have used $g\equiv g_{11}=g_{22}\ne g_{12}$. The chemical potential $\mu$ is found by solving $\int \rho(x) dx = N$. The spatial dependence of the density leads to the critical coupling strength $\Omega_{\rm cr}$ quoted in the main text being spatially dependent 
\begin{equation}\label{eq:omegaCr_x}
\hbar\Omega_{\rm cr}(x) = (g_{12}-g) \frac{\mu - V(x) + \hbar \Omega(0)}{g_{12}+g},
\end{equation}
and the miscibility condition is $\Omega(0)>\Omega_{\rm cr}(x)$ for all $x$. We have numerically checked the validity of this expression for our trapping parameters within the local denxity approximation.

\section{Control parameter}

The control parameter $\epsilon$ [see Eq. (2) of the main text] for a harmonically trapped BEC is spatially dependent due to the inhomogeneity of the critical coupling in Eq.~(\ref{eq:omegaCr_x})
\begin{eqnarray}\label{eq:trappedepsilon}
\epsilon_{\rm tr}(x,t) &=&  1 - \frac{\Omega(t)}{\Omega_{\rm cr}(x)},\nonumber\\
&=& 1-2\frac{\rho(0)}{\rho(x)}\left(1-\frac{t}{\tau_{\rm Q}}\right),
\end{eqnarray}
where $\rho(x)$ is given by Eq.~(\ref{eq:TFC}). The phase transition now has a travelling front, beginning at $x=0$  where the density is largest, and propagating outwards towards the edge of the BEC. The  position of the front as a function of time is found from the solution of $\epsilon_{\rm tr}(x,t)=0$ \cite{zureksoliton} which yields
\begin{equation}\label{eq:front}
x_{\rm F}(t) = \pm \sqrt{\frac{2}{m\omega^2}\left[\mu+\hbar\Omega(0)\right]\left(\frac{2t}{\tau_{\rm Q}}-1\right)}.
\end{equation}
It follows that the front moves with velocity
\begin{equation}\label{eq:v_front}
v_{\rm F}(x_{\rm F}) = \left|\frac{dx_{\rm F}}{dt}\right| = \frac{2[\mu+\hbar\Omega(0)]}{m\omega^2\tau_{\rm Q}|x_{\rm F}|},
\end{equation}
which diverges for $|x_{\rm F}| \rightarrow 0$.
For very fast quenches, $\tau_{\rm Q}\rightarrow 0$, the front is always faster than the speed of sound. For sufficiently long quenches there is a region $\tilde{X}$ where the speed of the front is smaller than the local speed of sound 
\begin{equation}
v_{s}(x) = \sqrt{\frac{(g + g_{12}) \rho(x)}{m}},
\end{equation} 
and domain formation is suppressed. The region $\tilde{X}$ grows in size as $\tau_{\mathrm{Q}}$ increases, hence reducing the length of the condensate where domains form according to the homogeneous KZ mechanism. The increase of the size of the suppression region with the quenching time contributes to the modified scaling exponent for spatially inhomogeneous phase transitions. For a homogenous system with $v_{\rm F} \ll v_{\rm s}$ the scaling exponent for the binary BEC is predicted to be $n=4/3$ \cite{FollowUpPRA}.  However, our result of $n\approx 1/2$ suggests that causality does not play a major role in the trapped system.
\bibliographystyle{apsrev4-1}